# Enhancing Road Safety: An IoT-Based Accident Detection and Prevention Mechanism

Prabhu Pugalenthi, MS Computer Science, University of Southern California, Los Angeles, CA
Pramod Krishnaa Dhanbalan, M.Sc. Software Systems, Coimbatore Institute of Technology, Tamil Nadu, India.

Abstract: Road traffic accidents remain a critical global crisis, consistently serving as a primary driver of preventable mortality and severe injury. These incidents are frequently precipitated by human error, including overspeeding, driving under the influence of alcohol, and cognitive fatigue. To address this urgent public safety challenge, this paper presents an intelligent, Internet of Things (IoT)-based Accident Prevention and Detection System (APDS) designed to systematically mitigate driver risk and optimize post-collision emergency responses. The proposed framework features a multi-tiered architecture capable of executing continuous real-time telemetry monitoring, proactive local alarm triggering, and automated situational intervention. Furthermore, the system integrates automated emergency communication protocols that aggregate immediate spatial coordinates via GPS and dispatch targeted alerts to medical facilities in close proximity, thereby optimizing response times and reducing accident-related fatalities.

## INTRODUCTION

The Internet of Things (IoT) comprises a network of physical entities equipped with sensors, software, and auxiliary technologies designed to facilitate data exchange with other devices and systems via the internet. The IoT market has demonstrated significant expansion; valued at USD 330.3 billion in 2021, it is projected to grow at a Compound Annual Growth Rate (CAGR) of 16.7% through 2026. This technological infrastructure has become an essential component of contemporary industrial and consumer frameworks.

### *Ubiquity of IoT Integration*

IoT systems permeate diverse sectors, including residential automation, industrial manufacturing, and healthcare. In residential contexts, sensor networks manage environmental controls and energy efficiency. Within industrial sectors, IoT facilitates production monitoring and logistics. A prominent application is the remote operation of machinery in hazardous environments, such as offshore oil rigs, where telemetry allows for shore-based oversight. Furthermore, supply chain management utilizes automated inventory tracking to optimize distribution channels.

In the medical domain, IoT-enabled devices, such as activity trackers and heart rate monitors, provide continuous patient telemetry. These applications illustrate that IoT serves as a fundamental pillar of the modern global infrastructure.

### *IoT in Automobile Industry*

The automotive industry leverages embedded sensor networks to monitor vehicular metrics, including spatial coordinates, velocity, and mechanical integrity. Contemporary vehicles increasingly rely on these intelligent systems to manage everything from tire pressure to engine thermal regulation. Despite these advancements, accident rates have escalated, predominantly due to human error. Statistics from the Ministry of Road Transport and Highways reveal that overspeeding accounted for 72.3% of road accidents and 71.2% of total deaths in 2022. Additional factors, such as driving under the influence and mobile device usage, contributed to 8.3% of fatalities. This research aims to augment existing vehicular systems with supplementary sensors designed to mitigate human error and optimize emergency response via automated proximity alerts and hospital notifications to aid the injured at the utmost time.

## GOAL AND OBJECTIVE

The primary aim of this project is to incorporate an IoT framework into the car's main Electronic Control Unit (ECU) by installing various sensors within the cabin to continuously monitor the vehicle's environment. This project focuses on understanding the functionality of these sensors to enhance driver safety and respond effectively to hazardous conditions.
A key component is the integration of an alcohol detection sensor that measures alcohol levels in the cabin before and during driving, ensuring the driver remains sober [5].
Additionally, a camera mounted on the dashboard will capture road speed limits, which will be processed using machine learning models on edge devices to alert the driver if they exceed the speed limit. Furthermore, a camera module is positioned in the driver's Heads-Up Display (HUD) to monitor driver drowsiness. This system will play an alert sound to notify the driver if signs of fatigue are detected, thereby promoting safer driving practices.
Accelerometer sensors will also monitor the car's resting angle and provide alerts if there are sudden changes in gradient [6], following guidelines from the Indian Road Congress. A gravity push button sensor will be pushed when the airbag is deployed during accidents, while a GPS sensor will monitor the vehicle's location and send emergency alerts to designated contacts in the event of an accident[5].
If the vehicle does not have built-in internet connectivity, a GSM module with a SIM card will facilitate emergency communication. Overall, this project aims to study and integrate these technologies to improve driver safety and automate real-time alerts during emergencies.

## MATERIALS AND METHODS

### *SYSTEM ARCHITECTURE AND SENSOR FUSION*

The proposed Smart Accident Detection and Response System (SADRS) utilizes a sensor fusion approach to provide comprehensive vehicular monitoring. The architecture integrates an MQ3 alcohol sensor for sobriety screening, an ADXL345 accelerometer for impact and orientation monitoring, and an OV7270 camera module for visual data processing. These hardware components are orchestrated by an Arduino UNO microcontroller, while a NodeMCU module facilitates cloud connectivity and real-time data transmission to the Blynk IoT platform. Emergency communications are handled by a SIM800L GSM module and a NEO 6M GPS unit to ensure precise location tracking during incident reporting.

*SYSTEM DESIGN AND ARCHITECTURE*

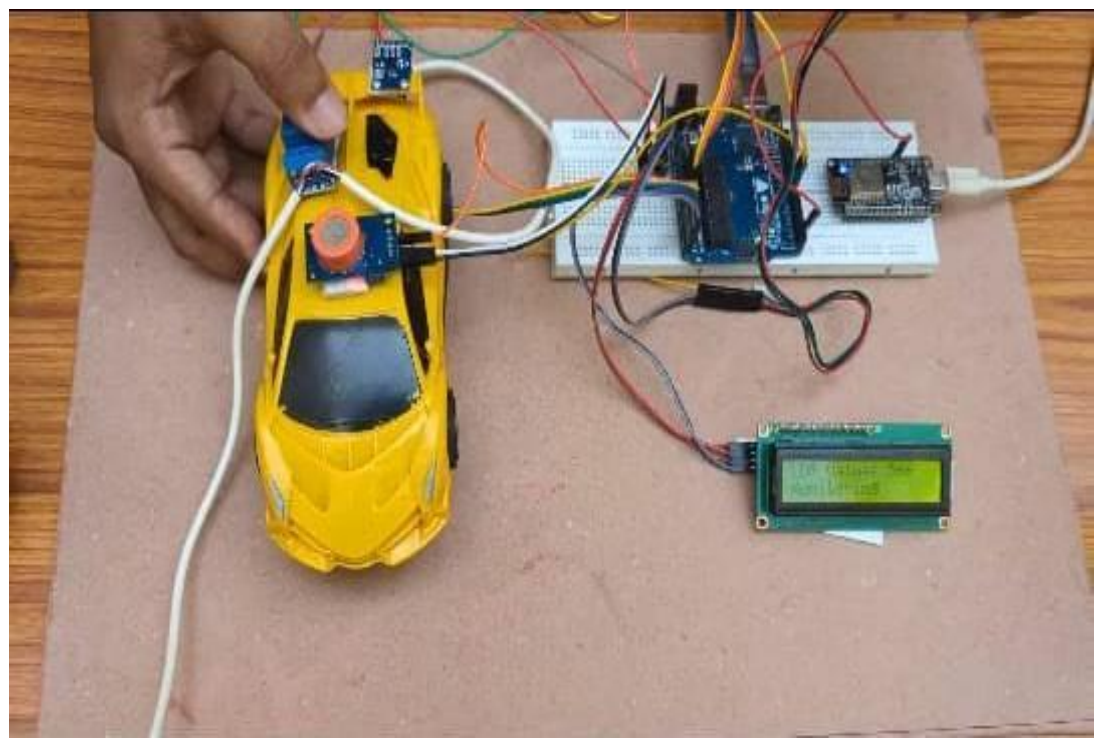

Figure 1: Miniature Model of SADRS

The proposed model requires careful monitoring when implemented in real-time. For our setup, we have developed a miniature version of the model (Figure 1) to demonstrate its functionality. The alcohol sensor should ideally be installed either on the dashboard or the roof inside the vehicle, positioned perpendicularly to the driver's seat. Installing the sensor on the dashboard above the steering wheel allows the driver to take a breath test before starting the engine. Upon entering the vehicle and turning on the ignition, an alert will prompt the driver to complete the alcohol breath test before the car can be started. The system will prevent the engine from starting until the driver completes the test and receives a negative result. Once the driver passes the alcohol test, they are permitted to drive. This continuous monitoring of alcohol levels within the cabin aims to significantly reduce accidents due to drunk driving.

Additionally, installing a camera module on either side of the front windshield will enable monitoring of speed limits and road signs by capturing images for processing through computer vision. Pre-trained machine learning models using Convolutional Neural Networks (CNN) will be employed to detect and extract relevant data from these images. This processing occurs within the car's system and its processing units. When speed limits or road signs are detected, alerts will be displayed on the dashboard to inform the driver. If the driver exceeds the speed limit, continuous beeping will serve as a reminder to slow down.

The Ministry of Road Transport and Highways (MoRTH) sets standard guidelines for road gradients. According to these guidelines, the ruling gradient typically ranges from 5% (1 in 20) for mountainous terrain to 6% (1 in 16.7) for steep terrain. The limiting gradient falls between 6% (1 in 16.7) and 7% (1 in 14.3), while the exceptional gradient, used in very challenging areas, ranges from 7% (1 in 14.3) to 8% (1 in 12.5) [7].

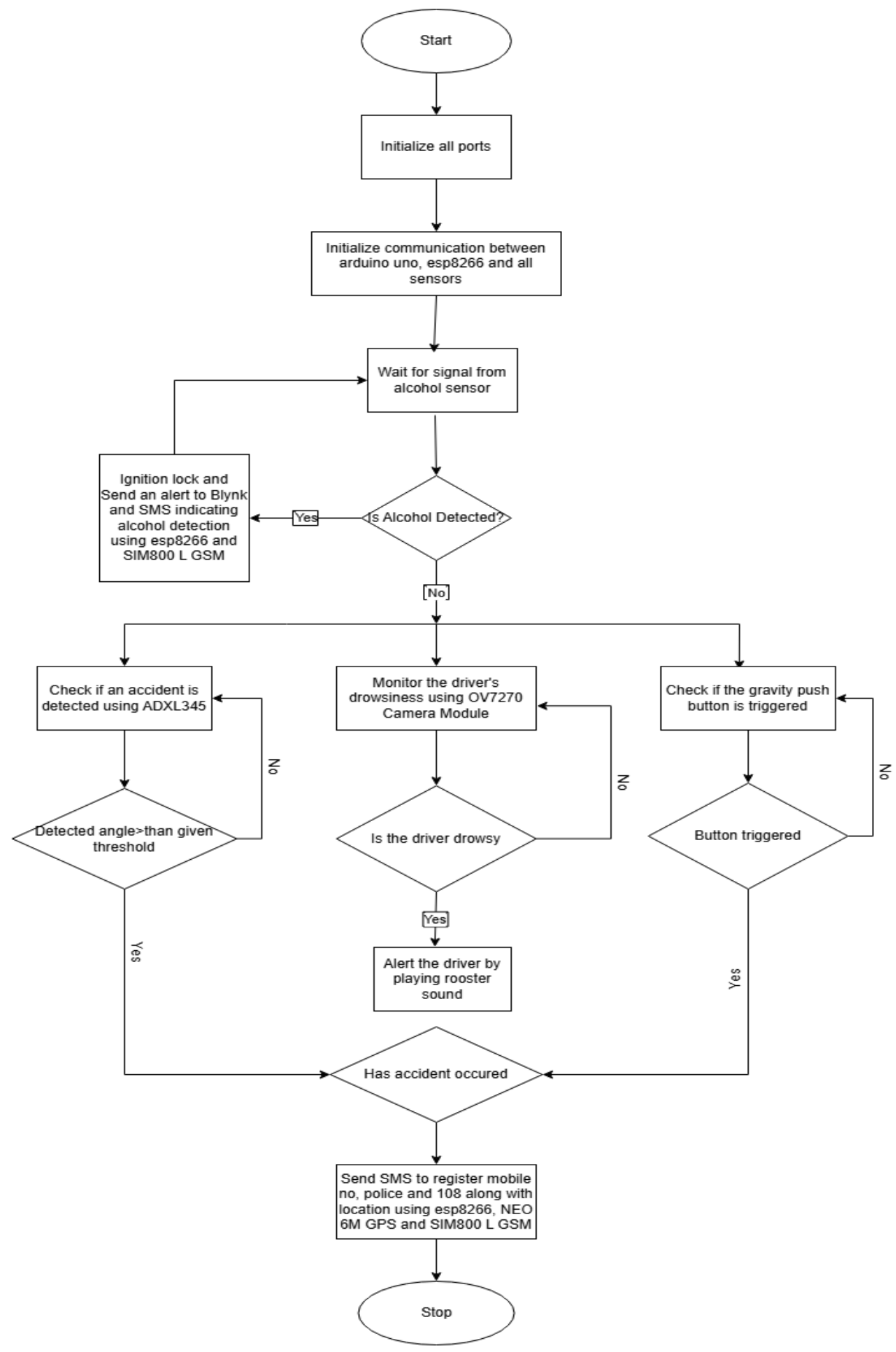

Figure 2: Control Flow of the system

In cases of head-on collisions, another sensor, a gravity pushes down sensor will be placed in front of the airbag system. When an airbag deploys due to such an impact, this sensor will activate when the airbag is pushed and when it hits the button, confirming that an accident has occurred. When an accident is detected, signals are sent to an Arduino UNO module that serves as the central controller. This signal is then transmitted to a GSM module, which alerts emergency contacts stored in the driver's mobile phone. Along with these alerts, the vehicle's location is shared with emergency contacts via GPS. The same GPS module is utilized to send location information to emergency services to dispatch an ambulance from a nearby station promptly, ensuring timely assistance at the accident scene [8]. Refer (Figure 2) for the control flow of the system and (Figure 3) for the circuit design of the system.

*IMPLEMENTATION*

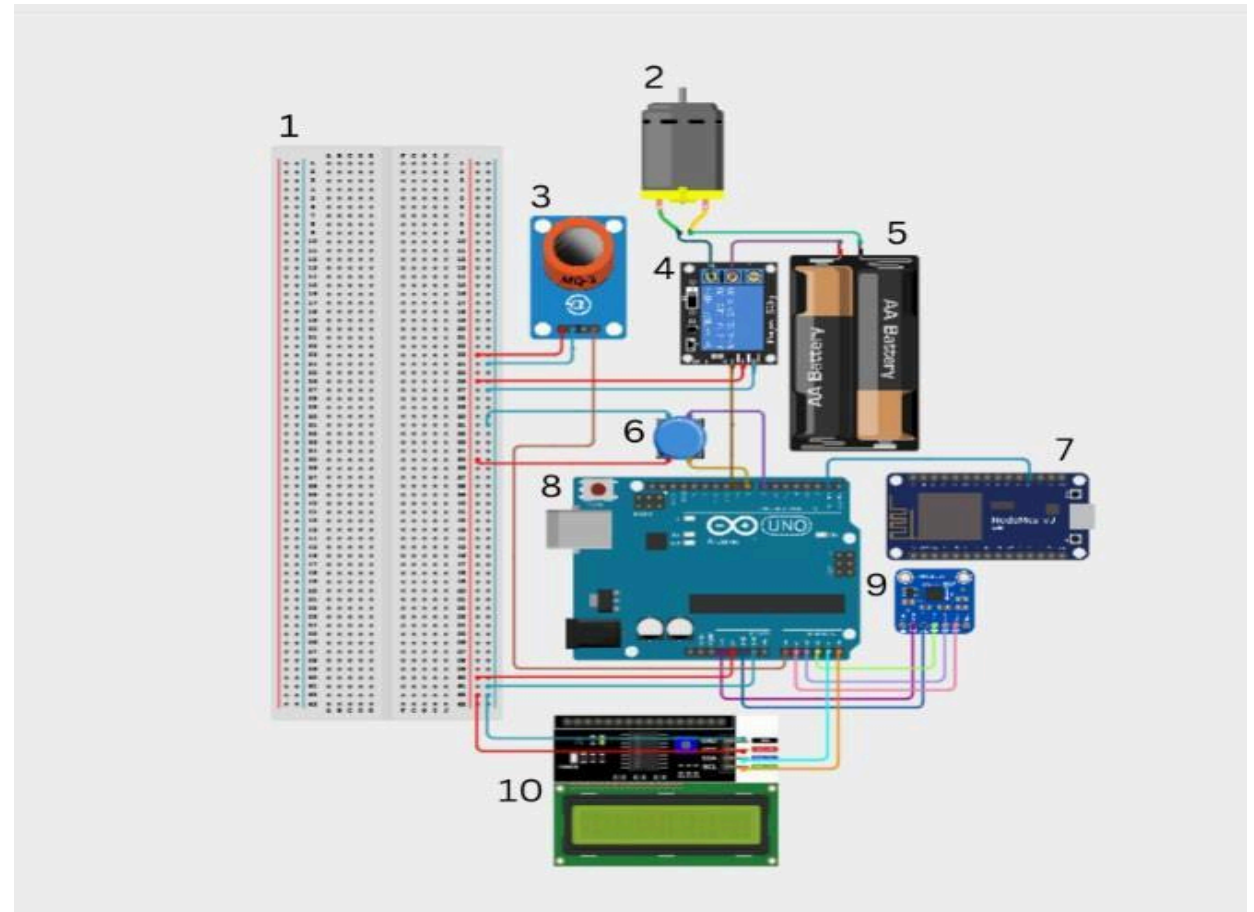

Figure 3: Circuit Diagram

1) Breadboard: Serves as the base for connecting multiple components without soldering, allowing easy testing and adjustments for circuit design.

2) DC Motor: Replaces the car engine in this model to simulate movement. It's used to indicate the starting mechanism of the vehicle, controlled by the system based on the alcohol sensor reading.
3) MQ3 Alcohol Sensor: Positioned near the steering wheel, this sensor prompts the driver to take an alcohol test before starting the vehicle. It detects alcohol levels in the air within a range of 25 ppm to 500 ppm.
4) Relay Module: Acts as a switch to control the power to the DC motor, allowing or preventing the vehicle from starting based on input from the alcohol sensor.
5) Battery Pack: Provides power to the motor and other components, simulating the car's battery in this model.
6) Push Button: Used to simulate ignition, allowing the driver to attempt starting the vehicle after taking the alcohol test. It triggers the system to check for alcohol levels before starting the motor.
7) NodeMCU: Manages communication between the sensors over a network, such as a mobile hotspot, enabling remote monitoring and control.
8) Arduino UNO: Serves as the main microcontroller, integrating all sensors and controlling their functions. The Arduino IDE is used to program it, enabling communication between components and processing of sensor data.
9) ADXL345 Accelerometer Sensor: Mounted under the vehicle's chassis to monitor changes in the car's angle. With a baseline set to 0 degrees and a threshold of 30 degrees, it triggers an alert whenever the car's angle exceeds this limit, indicating a potential accident.
10) LCD Display: Provides real-time feedback to the user, showing sensor readings and system status, including alcohol levels, angle, and pressure data.

*DRIVER DROWSINESS DETECTION*

Driver drowsiness is a major cause of road accidents, making its detection crucial for enhancing driver safety. This project introduces a Driver Drowsiness Detection system that uses computer vision and machine learning to monitor the driver's eye movements. By capturing real-time images of the driver's face, the system detects signs of tiredness or sleepiness based solely on eye behavior.

The system leverages a Convolutional Neural Network (CNN) trained on the UMass MRL Eye Dataset, which includes a variety of labeled eye images for identifying drowsiness indicators like prolonged eye closure[9]. Designed for both accuracy and speed, the model assesses the driver's alertness in real-time with minimal delay. By running on an embedded processing unit within the vehicle, this setup ensures low latency and reliable performance, providing a practical solution for promoting safer driving through early drowsiness detection[10].

*MODEL ACCURACY AND PERFORMANCE EVALUATION*

The driver drowsiness detection model has a high accuracy rate, which is crucial for real-time applications where quick and reliable alerts can help prevent accidents. It was trained and tested on a dataset with labeled images of drivers showing different levels of alertness and drowsiness. During testing, the model achieved an overall accuracy of about 91.9 % in detecting signs related to drowsiness. We also calculated performance metrics like precision, recall, and F1 score to evaluate how well the model balanced correctly identifying drowsiness signs (true positives) while avoiding misclassifications (false positives). These metrics show that the model is reliable in real-world driving situations, where both missing signs of drowsiness and incorrectly reporting them can affect safety. Refer (Table I) for the recorded Observations.

TABLE I. CNN MODEL ACCURACY PERCENTAGE

| Condition | Precision (%) | Recall (%) | F1-Score (%) | True Positives | False Positives | True Negatives | False Negatives |
|---|---|---|---|---|---|---|---|
| Alert | 95.2 | 93.7 | 94.4 | 250 | 15 | 240 | 17 |
| Slightly Drowsy | 90.4 | 88.6 | 89.5 | 190 | 20 | 230 | 25 |
| Very Drowsy | 92.1 | 91.5 | 91.8 | 220 | 18 | 210 | 22 |

**CALCULATING VALUES:**

a) Total True Positives (TTP) = 250+190+220 = 660
b) Total True Negatives (TTN) = 240+230+210 = 680
c) Total False Positives (TFP) = 15+20+18=53
d) Total False Negatives (TFN) = 17+25+22=64
e) Total Population (TP) =

TTP + TTN + TFP +TFN =660+680+53+64=1457

$$\text{Accuracy} = ((TP + TN) / (TP + TN + FP + FN)) \times 100$$

$$\text{Accuracy} = ((660 + 680) / 1457) \times 100 \approx 91.97\% \approx 92\%$$

Testing of the driver drowsiness detection model involved both pre-recorded video samples and real-time streaming to assess its accuracy in various lighting and environmental conditions. The model was evaluated on a validation set to ensure it could generalize beyond the training data. The system was also integrated with the miniature car for testing under different conditions, including daytime, nighttime, and artificial lighting. Overall, the model showed consistent high accuracy despite variations in external factors.

## RESULTS AND DISCUSSION

We implemented all sensors in a miniature car model, replacing the engine with a DC motor, and connected them to the Arduino main board. Accident-like scenarios were simulated, and we observed readings and alerts on the Blynk IoT application for registered emergency contacts via the Blynk Cloud app [11].

Figure 4 shows screenshots from the mobile application, including accelerometer readings in the XYZ plane, alcohol levels on the LDR scale, and downward force sensor status. A value of "0" indicates no force detected, while a value of "1" indicates airbag deployment. The screenshots also capture notifications for these alerts.

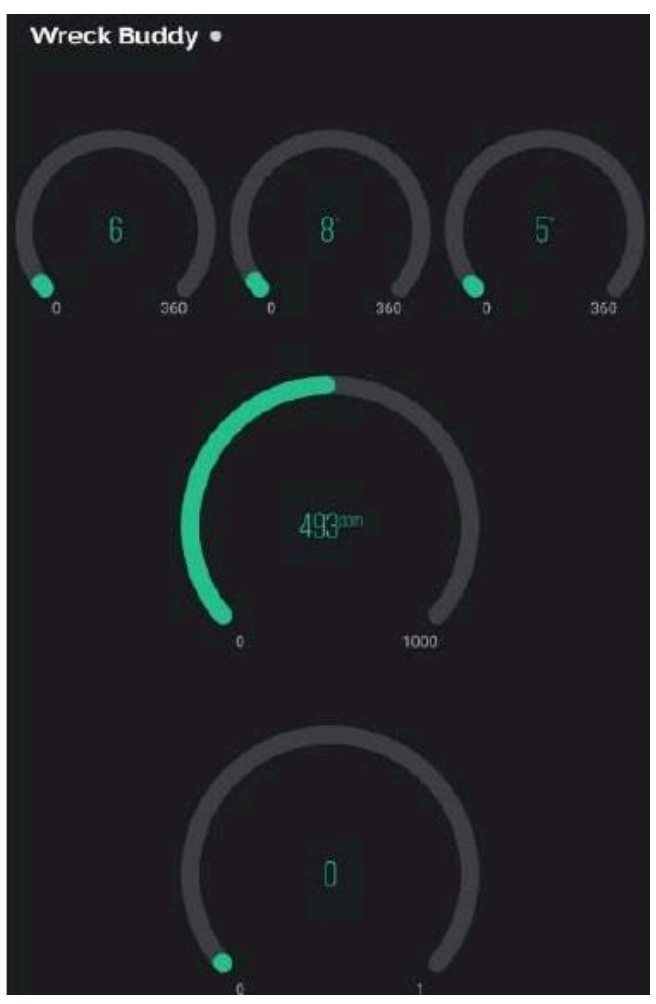

Figure 4: Blynk IoT readings and notifications triggered upon detection of anomalies or accident occurrence

## FUTURE ENHANCEMENTS

*A.* Adaptive Image Processing: Implementation of real-time algorithms that adjust image processing based on weather conditions, such as rain, fog, or nighttime. This approach will enhance image quality and improve the visibility of road signs, leading to better detection accuracy.

*B.* GPS Data Integration: Combining computer vision with GPS data to enhance the detection of road signs, speed limits, and other relevant information. This integration allows the system to tailor alerts based on geographic locations, such as school zones or construction areas.

*C.* Accident Prediction Models: Train machine learning models to analyze driving patterns like sudden steering changes or rapid speed variations to predict potential accidents. By providing early warnings, these models can help prevent accidents before they occur.

*D.* Multi-Modal Drowsiness Detection: Integration of facial recognition with additional sensors, such as heart rate monitors and EEGs, to improve drowsiness detection. A multi-modal approach will consider a wider range of indicators, making detection more reliable and accurate.

*E.* Cloud-Based Continuous Learning: Establishing a cloud-based system that allows the model to continuously learn from new data collected in real-world driving scenarios. This capability will help the model adapt to changing road conditions and driver behaviors over time.

*F.* Privacy Considerations: Addressing privacy concerns related to real-time facial recognition and monitoring driver behavior by offering options for anonymizing data or allowing drivers to control what information they share, particularly regarding facial images and health metrics.

*G.* Transparency in Data Usage: Ensuring transparency in data collection by informing drivers about what data is being collected and how it will be used. This practice builds trust and encourages drivers to adopt the technology.

By implementing these strategies, the driver drowsiness detection system can significantly improve its effectiveness in enhancing road safety while addressing privacy concerns and adapting to various driving condition

## CONCLUSION

In conclusion, this project aims to improve driver safety by integrating an IoT framework into the vehicle's main Electronic Control Unit (ECU). Equipped with multiple sensors within the cabin, the system continuously monitors the vehicle's surroundings and responds proactively to potential hazards. Key components include an alcohol detection sensor to ensure drivers are sober, a camera module to monitor speed limits, and a drowsiness detection system that warns drivers when signs of fatigue are detected. Additionally, an accelerometer will track the vehicle's resting angle, while a GPS module will provide real-time location data for emergency alerts. Together, these technologies not only work to prevent accidents but also enable automatic, real-time alerts during emergencies. By addressing critical safety concerns with advanced technology, this project has the potential to substantially reduce accidents and create a safer driving environment for everyone.

## ACKNOWLEDGMENTS

The authors wish to express their gratitude to the faculty and staff of the Coimbatore Institute of Technology for their technical guidance and for providing the resources necessary to conduct this research.

## REFERENCES

[1] "What is the Internet of Things (IoT)? | IBM." Accessed: Dec. 03, 2024. [Online]. Available: https://www.ibm.com/topics/internet-of-things
[2] "Internet of Things (IoT) Market Size, Statistics & Trends, Growth Analysis & Forecast, [Latest]," MarketsandMarkets. Accessed: Dec. 03, 2024. [Online]. Available: https://www.marketsandmarkets.com/Market-Reports/internet-of-things-market-573.html
[3] I. Limited, "IoT In Supply Chain Management | Infosys BPM." Accessed: Dec. 03, 2024. [Online]. Available: https://www.infosysbpm.com/blogs/supply-chain/internet-of-things-supply- chain.html
[4] "Road Accidents in India - 2022," MINISTRY OF ROAD TRANSPORT AND HIGHWAYS (TRANSPORT RESEARCH WING).[Online]. Available:https://morth.nic.in/sites/default/files/RA_2022_30_Oct.pdf
[5] S. K. R. Mallidi and V. V. Vineela, "IoT based smart vehicle monitoring system," Int. J. Adv. Res. Comput. Sci., vol. 9, no. 2, pp. 738– 741, 2018.
[6] S. Sharma and S. Sebastian, "IoT based car accident detection and notification algorithm for general road accidents," Int. J. Electr. Comput. Eng. IJECE, vol. 9, no. 5, p. 4020, Oct. 2019, doi: 10.11591/ijece.v9i5.pp4020-4026.
[7] Indian Roads Congress, IRC 052: Guidelines for the Alignment Survey and Geometric Design of Hill Roads (Third Revision). INDIAN ROADS CONGRESS, 2019. Accessed: Dec. 03, 2024. [Online]. Available: https://law.resource.org/pub/in/bis/irc/irc.gov.in.052.2019.pdf
[8] A. Shaik et al., "Smart car: An IoT based accident detection system," in 2018 IEEE global conference on internet of things (GCIoT), IEEE, 2018, pp. 1–5. Accessed: Dec. 03, 2024. [Online]. Available: https://ieeexplore.ieee.org/abstract/document/8620131/
[9] M. Kawulok, M. E. Celebi, and B. Smolka, Eds., Advances in Face Detection and Facial Image Analysis. Cham: Springer International Publishing, 2016. doi: 10.1007/978-3-319-25958-1.
[10] B. Fu, F. Boutros, C.-T. Lin, and N. Damer, "A Survey on Drowsiness Detection–Modern Applications and Methods," IEEE Trans. Intell. Veh., 2024, Accessed: Dec. 03, 2024. [Online]. Available: https://ieeexplore.ieee.org/abstract/document/10517310/
[11] "Get started with Blynk." Accessed: Dec. 03, 2024. [Online].Available: https://blynk.io/getting-started
[12] "What is Traffic Sign Recognition in Cars? - ADAS 101." Accessed: Dec. 03, 2024. [Online]. Available: https://caradas.com/understanding-adas-traffic-sign-recognition/